\documentclass{aa}
\usepackage{graphicx}
\def\hmpc{~h$^{-1}$ Mpc~}

\begin{document}
%
%
%
\title{ Abell 3560, a galaxy cluster at the edge of a major merging event } 
%
%
\author
{
S.~Bardelli\inst{1} \and
T.~Venturi\inst{2} \and
E.~Zucca\inst{1} \and
S.~De Grandi\inst{3} \and
S.~Ettori\inst{4} \and
S.~Molendi\inst{5}
}

\institute
{
INAF - Osservatorio Astronomico di Bologna, 
via Ranzani 1, I--40127 Bologna, Italy
\and
Istituto di Radioastronomia del CNR, via Gobetti 101, I-40129, Bologna, Italy 
\and
INAF - Osservatorio Astronomico di Brera, Via Bianchi 46, I-23807 Merate
(LC), Italy 
\and
European Southern Observatory, Karl-Schwartzschild-Strasse 2, D-85748 Garching,
Germany
\and
Istituto di Fisica Cosmica ``G.Occhialini'' del CNR, Via Bassini 15, I-20133
Milano, Italy 
}
%
\date{Received 17 May 2002 / Accepted 28 August 2002}
%
\titlerunning{Abell 3560}
\authorrunning{S. Bardelli et al.}
\abstract{
In this paper we study A3560, a rich cluster 
at the southern periphery of the A3558 complex, a chain of interacting
clusters in the central part of the Shapley Concentration supercluster.
\\
From a ROSAT-PSPC map we find that the X-ray surface brightness 
distribution of A3560 is well 
described by two components, an elliptical King law and a more peaked and 
fainter structure, which has been modeled with a Gaussian. 
The main component, corresponding to the cluster, is elongated with the
major axis pointing toward the A3558 complex. The second component,
centered on the Dumb-bell galaxy which dominates the cluster, appears
significantly offset (by $\sim 0.15$ \hmpc) from the cluster X-ray 
centroid. 
\\
From a Beppo-SAX observation we derive the radial temperature profile,
finding that the temperature is constant (at $kT \sim 3.7$ keV) up to
8 arcmin, corresponding to 0.3 \hmpc: for larger distances, the temperature
significantly drops to $kT \sim 1.7$ keV.
We analyze also temperature maps, dividing the cluster into 4 sectors and
deriving the temperature profiles in each sector: we find that the
temperature drop is more sudden in the sectors which point towards the A3558 
complex. 
\\
From VLA radio data, at 20 and 6 cm, we find a peculiar bright extended
radio source (J1332-3308), composed of a core (centered on the northern 
component of the Dumb-bell galaxy), two lobes, a ``filament" and a diffuse
component. The morphology of the source could be interpreted either by 
a strong interaction of the radio source with the intracluster medium
or by the model of intermittency of the central engine.
\keywords{
X-rays: galaxies: clusters -
galaxies: clusters: general - 
galaxies: clusters: individual: A3560
}
}
\maketitle

%
%
\section{Introduction}

\begin{figure}
\centering
\includegraphics[angle=0,width=\hsize]{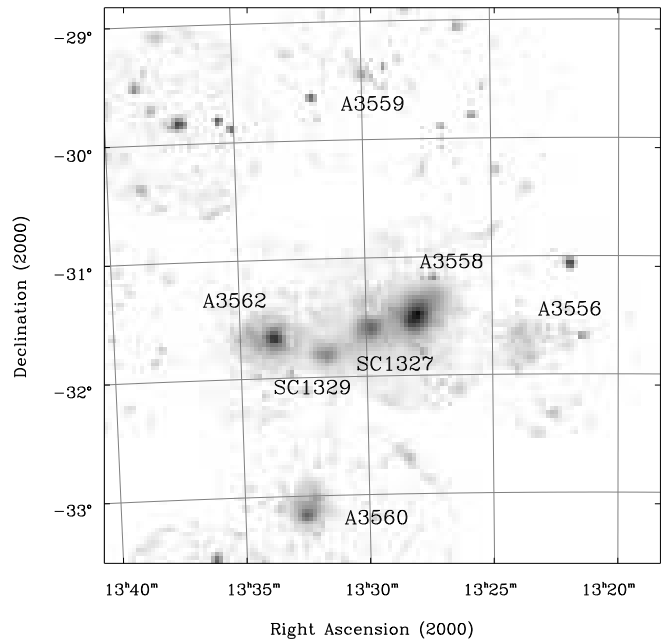}
\caption{Large scale view of the center of the Shapley Concentration centered 
on the A3558 complex. This image is a mosaic of all the pointed ROSAT-PSPC  
frames available in this region (see Ettori et al. \cite{ettori97}). 
Note the position of the cluster A3560 (on the bottom). 
}
\label{fig:largeview}
\end{figure}

Merging has been recognized as the leading process in massive cluster
formation, as a consequence of the hierarchical structure formation 
scenario. A large amount of numerical work has been done to study this
phenomenon at all the relevant scales: among others,
Ricker et al. (\cite{ricker01}) studied in detail the physics of the
plasma during the collision of two clusters under different initial conditions,
while Colberg et al. (\cite{colberg99}) analyzed the role of the
cosmological environment on the merging.
\\ 
From the observational point of view, the improvements of the 
Point Spread Function and sensitivity of the Chandra satellite
led to the detailed description of the shocks (Markevitch et al. 
\cite{markevitch02}) and the discovery of the so-called ``cold fronts" 
(Vikhlinin et al. \cite{vikhlinin01}), which are direct consequences of 
merging at an advanced state.
\\
Little work has been done on the global, multiscale  description of 
this phenomenon. To this end, we are carrying on a long term project 
aimed at studying the merging in the particularly rich environment of the 
central part of the Shapley Concentration supercluster. 
In this region, three ``cluster complexes" are found (Zucca et al. 
\cite{zucca93}), i.e. structures of $\sim 7$ \hmpc (hereafter $h=H_o/100$) 
which represent major cluster mergings at various evolutionary stages.   
\\  
The most massive structure, the A3558 complex (Figure \ref{fig:largeview}), 
is probably a collision seen
after the first core-core encounter (Bardelli et al. \cite{bardelli98b}).
The whole complex, formed by a chain of three ACO clusters and
two poor groups, is embedded in a hot gas filament (Bardelli et al.
\cite{bardelli96}; Kull \& B\"ohringer \cite{kull99}) and in a common 
envelope of galaxies (Bardelli et al. \cite{bardelli94}, \cite{bardelli98a}). 
The estimated mass ranges between $10^{15}$ and $10^{16} M_{\odot}$
(Bardelli et al. \cite{bardelli00}, Ettori et al. \cite{ettori97},
Reisenegger et al. \cite{reisenegger00}).
\\
The entire structure presents a large number of substructures, some of
which have an X-ray counterpart as diffuse emission (Bardelli et al. 
\cite{bardelli02}). 
Moreover, the merging seems to lead to a lack of radio sources with respect
to ``normal" clusters (Venturi et al. \cite{venturi00}).
The presence of a halo radio source, of a minihalo and a relic radio source
(Venturi et al., in preparation) is further evidence of ``stormy weather".
\\
In this paper we concentrate on the cluster A3560, a rich cluster at the
southern periphery of the A3558 complex. In Figure \ref{fig:largeview} 
a mosaic of the available ROSAT-PSPC frames is shown and gives the 
large-scale distribution of the clusters in this region. The two groups 
labelled SC1327 and SC1329 are SC$1327-312$ and SC$1329-313$.
\\   
The distance of A3560 from the nearest X-ray clump of the A3558 complex 
(corresponding to SC$1329-313$) is $\sim 3$ \hmpc. 
Given the proximity of such a large mass concentration, the high predicted 
infall velocity ($\sim 2000$ km s$^{-1}$, Reisenegger et al. 
\cite{reisenegger00}) and the existence of the underlying overdensity of the 
supercluster (see Bardelli et al. \cite{bardelli00}), a certain degree of
disturbance can be expected for A3560.    
\\
The plan of the paper is the following. In Sect.2 we describe the general
properties of A3560 and in Sect.3 we perform the spatial analysis of the 
ROSAT-PSPC map. In Sect.4 we analyze our new Beppo-SAX observations on this
cluster, obtaining temperature profiles and maps, while in Sect.5 we
present the properties of the central radio source of A3560.
Finally in Sect.6 we discuss and summarize the results.

%
%
\section{The cluster A3560}

Abell 3560 is a cluster of richness class 3 and Bautz-Morgan class I; 
the center has coordinates $\alpha_{2000}=13^h 32^m 22^s$, 
$\delta_{2000}=-33^o 05' 24''$. 
Note that the position reported by Abell, Corwin \& Olowin (\cite{aco}), 
i.e. $\alpha_{2000}=13^h 31^m 50^s$ and $\delta_{2000}=-33^o 13.4'$, 
corresponds neither to an optical overdensity nor to a diffuse X-ray emission.
Probably, there was a mistake in reporting the position, considering that 
nearby this position a bright galaxy is located (NGC5193): given its
redshift ($z\sim 0.012$), this galaxy is not associated with the cluster.
More details about this discrepancy can be found in Willmer et al. 
(\cite{willmer99}) and references therein.
\\
On the basis of 32 redshifts, Willmer et al. (\cite{willmer99}) 
estimated $\langle v \rangle =14470\pm 123$ km s$^{-1}$ and 
$\sigma=614 \pm 68$ km s$^{-1}$ and detected a marginal significance of 
substructure by applying the Lee 2-3D statistics. 
\\
This cluster is dominated by a Dumb-bell galaxy: Willmer et al. 
(\cite{willmer99}) reported the velocities of the two components,
consistent each other within the errors and at rest with the cluster 
velocity centroid.      
\\
At the redshift of the cluster ($z=0.048$), assuming $q_o=0.5$, 
1 arcmin corresponds to $\sim 38.56$ h$^{-1}$ kpc.

\begin{figure}
\centering
\includegraphics[angle=0,width=\hsize]{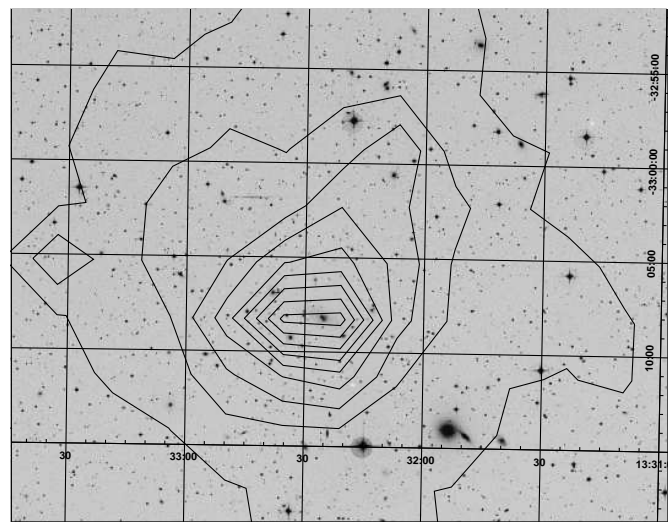}
\caption{ 
X-ray isophotes, from a ROSAT-PSPC observation pointed on A3560, superimposed
on the optical image from the Digital Sky Survey. The bright elliptical galaxy
(with a nearby edge-on spiral) in the lower right corner is NGC5193.
See the text for further details about the isophotes.
}
\label{fig:dssrosat}
\end{figure}
%
%
\section{ROSAT spatial analysis}

ROSAT-PSPC data for A3560 have been taken from the public {\it HEASARC} 
archive, and are part of the $\sim 6000$ seconds observation referenced as 
RP800381A02. 
In Figure \ref{fig:dssrosat} we present the X-ray isophotes superimposed
on the optical image from the Digital Sky Survey: the contours refer 
to the inner $20'$ of the ROSAT-PSPC data, where the vignetting is small.
The data have been smoothed with a Gaussian of 10 pixel FWHM 
(1 pixel $=$ 15 arcsec) and 
the linear step between contours is 0.30, with the lowest
isophote corresponding to 0.282 cts pix$^{-1}$ (well above the background
value, see below).
Already at a first look at the image, 
it is clear that the cluster is formed by two components 
with offset centers: one symmetric, circular component and one elongated,
more extended contribution.   
\\
For this reason, following Bardelli et al. (\cite{bardelli96}), we fitted the 
surface brightness distribution using an elliptical King law and a Gaussian 
of the form

\begin{eqnarray} 
P(x,y) &=& I_o \left[1+\left({{x'}\over{R_1}} \right)^2 + 
           \left({{y'}\over{R_2}}\right)^2 \right]^{-3\beta + 0.5} +  
                                                                \nonumber\\
       & & I_{g} {\rm exp}\left[ {{(x-x_g)^2+(y-y_g)^2}\over {2\sigma^2}}
           \right] + bck 
\end{eqnarray}
\\
where 

\begin{eqnarray*}
 x' &=& (x-x_k) \cos \theta + (y-y_k) \sin \theta    \\
 y' &=& - (x-x_k) \sin \theta + (y-y_k) \cos \theta 
\end{eqnarray*}
\\
in order to take into account the position angle $\theta$ of the cluster.
The variables to be estimated are the two normalizations $I_o$ and $I_g$,
the positions ($x_k$, $y_k$) and ($x_g$, $y_g$) of the centers of the
cluster and of the Gaussian respectively, the position angle of the cluster 
($\theta$),
its core radii ($R_1$ and $R_2$), the exponent of the King law ($\beta$), 
the width of the Gaussian ($\sigma$).
The background ($bck$) has been estimated by averaging the count-rates in
9 areas in the external parts of the map, and resulted to be 
0.0781  $\pm$ 0.0029 cts pix$^{-1}$.   
\\
\begin{figure}
\centering
\includegraphics[angle=0,width=\hsize]{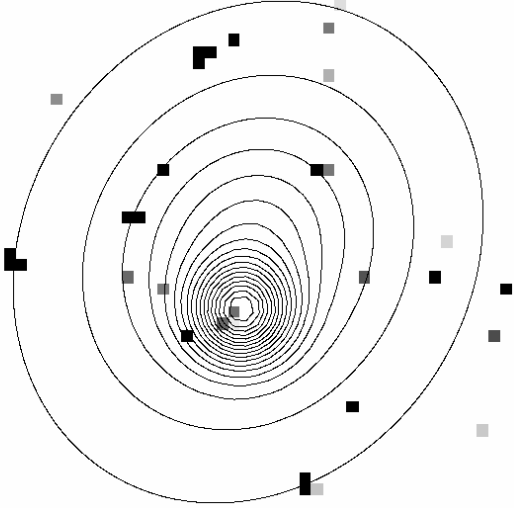}
\caption{ 
Model isodensity contours superimposed to a $\chi^{2}$ map of the difference
model-data. The pixels have a $3\times 3$ rebinning with respect to the original
frame. Visible points correspond to $\chi^2>6$ and have been clipped 
from the fit.
}
\label{fig:chi2fit}
\end{figure}
\begin{table*}
\caption[]{Results of the bi-dimensional fit (errors are at $1\sigma$ level).}
\begin{flushleft}
\begin{tabular}{lllll}
\hline\noalign{\smallskip}
Component & Normalization  & core radii & $\beta$ & $\theta$ \\
\noalign{\smallskip}
\hline\noalign{\smallskip}
King  & $0.93 \pm 0.05$ cts pix$^{-1}$ 
      & $33.75 \pm 1.63$ - $28.86 \pm 1.44$ pix & $0.525 \pm 0.022$ 
      & $32^o \pm 5^o$ \\
~~~~  & $3.20 \times 10^{-7}$ erg str$^{-1}$ cm$^{-2}$ s$^{-1}$ 
      & 0.325 - 0.278 h$^{-1}$ Mpc 
      & ~~ & ~~  \\
\noalign{\smallskip}
\hline\noalign{\smallskip}
Component & Normalization  & $\sigma$ & ~~ & ~~  \\
\noalign{\smallskip}
\hline\noalign{\smallskip}
Gaussian  & $1.69 \pm 0.38$ cts pix$^{-1}$ & $8.08 \pm 0.42$ pix &~~~ &~~  \\
~~~~~~~~  & $6.07 \times 10^{-7}$ erg str$^{-1}$ cm$^{-2}$  s$^{-1}$  
          & 0.08  h$^{-1}$ Mpc &~~~  &~~\\
\noalign{\smallskip}
\hline
\end{tabular}
\end{flushleft}
\label{tab:spatial}
\end{table*}
The fit has been performed by minimizing the $\chi^2$ variable between 
the model and the data, after having rebinned (3$\times$ 3 pixels) the 
original image. A number of strongly ($\chi^2>6$) deviant pixels, which
likely  correspond to real sources, have been
eliminated from the fit procedure.   
The results of the fit, done in the [0.5-2.0] keV band, are reported in    
Table \ref{tab:spatial}.
\\
In Figure \ref{fig:chi2fit} we show the model isodensity contours superimposed
to a $\chi^{2}$ map of the difference model-data. The pixels have 
a $3\times 3$ rebinning with respect to the original frame. 
\\
The  centers of the two resulted to be separated by
15.39 pixels ($3.8$ arcmin), corresponding to 0.15 \hmpc. 
In order to verify if the peaked component is really a Gaussian, 
we subtracted  the King model from the data map and analyzed the residuals
separately with various image fitting packages. In all cases the Gaussian model 
was compatible with the data and the dispersion parameter consistent with 
our ones. 
\\
The position angle of the King model is tilted $32$ degrees 
westward with respect to the North direction: note that this value is close 
to the direction toward  A3558 ($\sim 30$ degrees), which is $\sim 4.6$ \hmpc
away from A3560 and is thought to be the barycenter of the A3558 cluster 
complex.

%
\section{Beppo-SAX spectral analysis}

%
\subsection{Observations and data reduction}

The cluster A3560 was observed by the Beppo-SAX
satellite (Boella et al. \cite{boella97a}) in the period 1-3 July 2000,  
with a total exposure time of $65.5$ ksec.
\\
Here we discuss  the data from two of the instruments onboard Beppo-SAX: the 
Medium-Energy Concentrator Spectrometer (MECS) and the Low-Energy Concentrator
Spectrometer (LECS). The MECS (Boella et al. \cite{boella97b}) is  
composed of two units, working in the [1--10] keV energy range. At
6 keV, the energy resolution is $\sim 8\%$ and the angular resolution
is $\sim 0.7'$ (FWHM). The LECS (Parmar et al. \cite{parmar97}),
consists of an imaging X-ray detector, working in the [0.1--9] keV
energy range, with 20$\%$ spectral resolution and $0.8'$
(FWHM) angular resolution (both computed at 1 keV). Standard
reduction procedures and screening criteria have been adopted to
produce linearized and equalized event files. The MECS (LECS) data
preparation and linearization was performed using the {\sc Saxdas}
({\sc Saxledas}) package under {\sc Ftools} environment.
\\
We have taken into account the PSF-induced spectral distortions 
(D'Acri et al. \cite{dacri98}) in the MECS analysis using effective 
area files produced with the {\it effarea} program.
All MECS and LECS spectra have been background subtracted using spectra
extracted from blank sky event files in the same region of the
detector as the source (see Fiore et al. \cite{fiore99}).
A detailed explanation of the MECS analysis is given in De Grandi \& Molendi 
(\cite{degrandi01}).
\\
As done in Ettori et al. (\cite{ettori00}), for the LECS we have used two
redistribution matrices and ancillary response files, the first computed
for an on-axis pointlike source and the second for a source with a
flat brightness profile. The temperatures and abundances we derive in
the two cases do not differ significantly, as the telescope vignetting
in the [0.1--4.0] keV energy range is not strongly dependent upon energy.
All spectral fits have been performed using XSPEC Ver. 10.00,
fitting the data with a {\it mekal} (Mewe et al. \cite{mewe95},
Kaastra \cite{kaastra92}) model, absorbed for the
nominal Galactic hydrogen column density ({\it wabs} model). 
The hydrogen column has been fixed to the value of 
$4.1 \times 10^{20}$ cm$^{-2}$ (Dickey \& Lockman \cite{dickey90}).   

\begin{figure}
\centering
\includegraphics[angle=-90,width=\hsize]{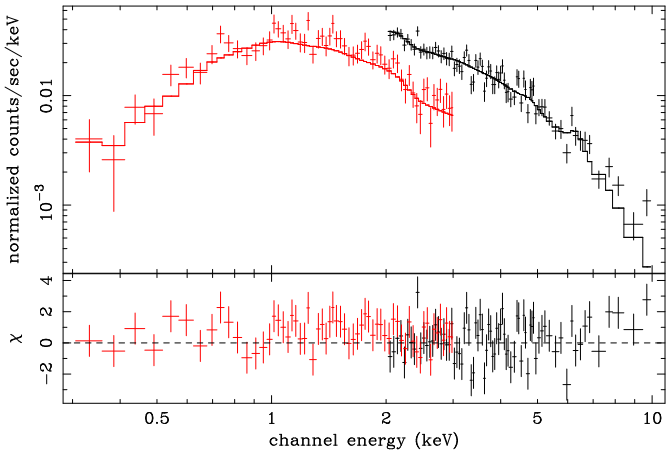} 
\caption{ 
Spectrum and fit of the MECS$+$LECS data within 8 arcmin 
from the center of A3560.
} 
\label{fig:lecsmecs}
\end{figure}

%
\subsection{Global temperature}

We fitted the spectrum of A3560 within 8 arcmin (correspondig to $\sim 0.3$ 
\hmpc).
First we estimated the temperature and abundance from the two combined MECS 
instruments, in the [2--10] keV range.
We found $kT=3.80^{+0.37}_{-0.32}$ keV and an abundance $0.10^{+0.10}_{-0.10}$,
where the errors are at the $90\%$ significance level. 
The reduced $\chi^{2}$ is $1.44$ with 109 degrees of freedom.
\\    
The temperature and abundance determination for the LECS in the range 
[0.3--3] keV gives
$kT=3.34^{+0.99}_{-0.66}$ keV and abundance $0.33^{+0.73}_{-0.33}$,  
with 1.08 of reduced $\chi^{2}$ and 251 degrees of freedom.
\\
Given the fact that these estimates are consistent with each other,
in order to enhance the statistics we fitted the combined LECS$+$MECS data.
First, we ran the fit by estimating also the relative normalization between 
the two instruments: having checked that this number is consistent (within 
one sigma) with the standard value of 0.5, we fixed it at 0.5. 
The results (see Figure \ref{fig:lecsmecs}) are $kT=3.69^{+0.24}_{-0.22}$ keV 
and abundance 
$0.13^{+0.10}_{-0.09}$ with a reduced $\chi^{2}$ of 1.19 and
363 degrees of freedom.  
\\
The temperature is in agreement with the value of 3.4 keV estimated by
Ebeling et al. (\cite{ebeling96}) on the basis of the $L_X-T$ relation.  
\\
Following the $\sigma-T$ relation of Lubin \& Bahcall (\cite{lubin93})
[$\sigma=332(kT)^{0.6}$ km s$^{-1}$], the determined temperature
implies a velocity dispersion of $727^{+17}_{-17}$ km s$^{-1}$ (one sigma
errors), which is consistent at 1.6 sigma with the value estimated
by Willmer et al. (\cite{willmer99}). 
\\
With the estimated temperature, the total luminosity within 16 arcmin
is $L_{[2-10]keV}=3.6 \times 10^{43}$ h$^{-2}$ erg s$^{-1}$ corresponding to a 
bolometric luminosity of  $L_X=7.1 \times 10^{43}$ h$^{-2}$ erg s$^{-1}$
well consistent with the PSPC-ROSAT value of  
$L_X=6.9 \times 10^{43}$ h$^{-2}$ erg s$^{-1}$ estimated by 
David et al. (\cite{david99}).

\begin{figure}
\centering
\includegraphics[angle=0,width=\hsize]{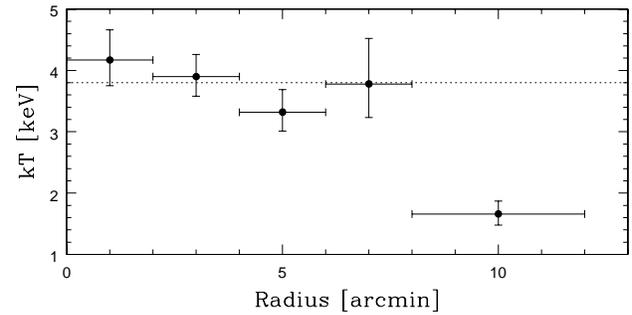} 
\caption{ 
Temperature radial profile of A3560. The vertical bars correspond to $68\%$
errors, while the horizontal bars represent the bins used to extract the
counts. The dotted line corresponds to the global fit, derived within 
8 arcmin.  
} 
\label{fig:temp_prof}
\end{figure}

\begin{figure}
\centering
\includegraphics[angle=-90,width=\hsize]{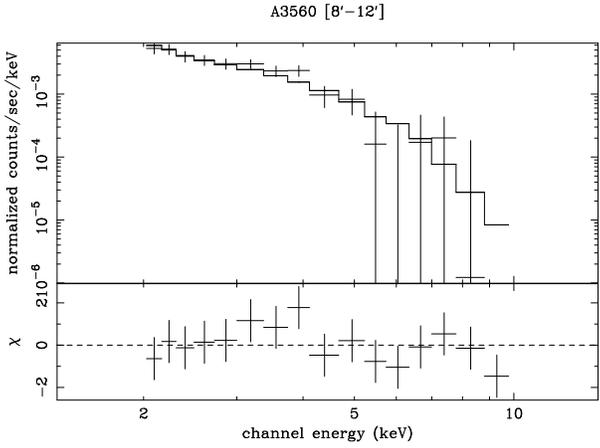} 
\caption{ 
Spectrum, fit and residuals of the MECS data in the 8-12 arcmin annulus.
} 
\label{fig:fit8-12}
\end{figure}

\begin{figure}
\centering
\includegraphics[angle=0,width=\hsize]{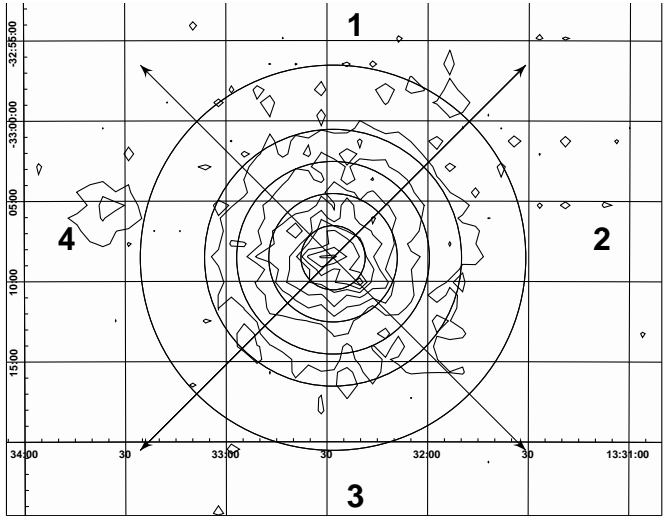} 
\caption{Beppo-SAX MECS image of A3560. The image has been smoothed with a 
Gaussian of 6 pixels FWHM, corresponding to $0.8$ arcmin. 
The concentric circles 
correspond to the bins of the radial profile, while the quadrants correspond
to the sectors used for the temperature map analysis. In this case,
the third and fourth shells have been used as a single radial bin.   
} 
\label{fig:mecs+grid}
\end{figure}

\begin{figure}
\centering
\includegraphics[angle=0,width=\hsize]{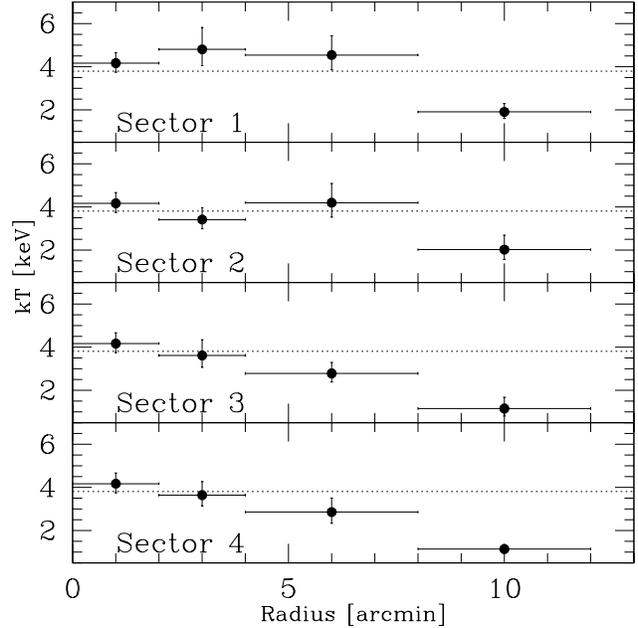} 
\caption{ 
Temperature map of A3560. The vertical bars correspond to $68\%$
errors, while the horizontal bars represent the bins used to extract the
counts. Dotted lines are drawn in correspondence of the global fit 
temperature. 
} 
\label{fig:temp_map}
\end{figure}

%
\subsection{Temperature profiles and maps}

In analysing temperature profiles and maps we used only the MECS data,
for which the correction for the PSF-induced spectral distortions is
available.
\\
The cluster emission has been divided into concentric annuli, centered on the 
X-ray emission peak: out to $8$ arcmin the annuli are $2'$ wide, beyond this
radius the annuli are $4'$ wide.
To all spectra accumulated from these annular regions, we have applied
the model described in Sect.4.1 to derive temperature and metal abundances. 
The fitting procedure stops at the last annulus where the source 
counts are more than $30\%$ of the total (i.e. source plus background) counts
(see the discussion in De Grandi \& Molendi \cite{degrandi02}).
\\
In Figure \ref{fig:temp_prof} we report the temperature profile of A3560
in annuli around the cluster center. The vertical bars correspond to the 
$68\%$ errors and the horizontal bars represent the bins used to extract the 
counts. The dotted line corresponds to the value obtained from the global fit.
\\
Regarding the abundance, the values derived in the first three bins 
are consistent with the global determination, while the fit was unconstrained 
for the last two points.
Therefore for radii larger than 6 arcmin the plotted temperatures were 
derived fixing the abundance to the global value.
\\
The  profiles are in agreement with those obtained by Bonamente et al. 
(\cite{bonamente01}) using ROSAT-PSPC data. 
In particular, it is encouraging that at 
large distances  from the center ($\sim 10$ arcmin), they show the same 
low temperature: given the fact that this is an independent estimate, 
the result is not an artifact of the MECS or of our background subtraction
procedure. 
The spectral fit of the data in the last annulus is shown in 
Figure \ref{fig:fit8-12}: the temperature is $1.65^{+0.37}_{-0.30}$ keV and
the reduced $\chi^{2}$ is $1.03$ with 62 degrees of freedom.
We note however that the robustness of the MECS temperature estimate at such 
distances was already demonstrated by Bardelli et al. (\cite{bardelli02}): 
with the use of two different overlapping exposures, we measured the 
temperature of the same point once with the inner part of the MECS and once 
with the external part, finding the same result. 
\\
The first point of Bonamente et al. (\cite{bonamente01}), corresponding to our 
first two radial bins, although at lower temperature ($2.2$ keV), 
is still consistent with our determinations. However, due to the different 
energy range, the PSPC is more sensitive to lower temperature than the MECS 
and therefore we cannot rule out the existence of a multiphase medium.  
\\
In order to explore if there is an asymmetry of the temperature distribution
due to the interaction of the intracluster medium with possible 
material of the A3558 complex, we divided the cluster map in four sectors 
as shown in Figure \ref{fig:mecs+grid}. Sector 1 is North of the image
and the numbers increase clockwise. In order to increase the statistics,
we used the third and fourth shells as a single radial bin.
In Figure \ref{fig:temp_map} the temperature profiles in the four 
sectors are presented.
In all the quadrants the last point is consistent to be at the same low
temperature, 
but the temperature drop is more sudden in sectors 1 and 2, while it is
smoother in the other sectors. Indeed, the temperature derived in the
$4'-8'$ annulus from the combined fit of sectors $1+2$ is 
$4.34^{+1.10}_{-0.75}$ keV, while for sectors $3+4$ it is 
$2.80^{+0.71}_{-0.47}$ keV: therefore the values are different at $\sim 2.5
\sigma$ confidence level.  
\\
Considering the region where the cluster is isothermal, at the global 
temperature of $kT=3.69$ keV, the derived total mass is 
$M(<0.3$\hmpc$) = (3.08 \pm 0.19) \times 10^{13}$ h$^{-1}$ $M_{\odot}$. 
If we extrapolate this mass up to a radius of 1 \hmpc, 
we find $M(<1$\hmpc$) = (1.89 \pm 0.12) \times 10^{14}$ h$^{-1}$ $M_{\odot}$,
consistent within the errors with the value of $1.64^{+0.93}_{-0.77} \times 
10^{14} M_{\odot}$ given by Willmer et al. (\cite{willmer99}), based on 
optical data within 30 arcmin (corresponding to $\sim 1.16$ \hmpc).
However, given the fact that this cluster presents a temperature gradient,
this extrapolation cannot be considered reliable. 
We tried to model the gradient in the gas temperature with a polytropic
profile (Ettori \cite{ettori2000}), finding an index $\gamma \ge 2$:  
this indicates that the intracluster medium is unstable to convective
mixing, with turbulent motions that relax to an homogeneous gas after
several sound crossing times [$t_s$(10 arcmin) $\approx 1.5 \times 10^9$ 
h$^{-1}$ yr $\approx 0.25\ t_{age\ Universe}$]. 
This fact questions the validity of the hydrostatic equilibrium hypothesis
for regions at distances $> 0.3$ \hmpc from the cluster center.
\\
The situation of the Gaussian component is also complex, because of
the lack of temperature determination. The MECS and LECS instruments
do not have enough spatial resolution to separate the spectra of the
Gaussian component from the King model; moreover, the luminosity of the
Gaussian component is only $\sim 2\%$ of the total luminosity. 
The best direct measurement can
be considered the central point of Bonamente et al. (\cite{bonamente01}) 
at $kT=2.2^{+1.5}_{-0.6}$ keV, obtained from a ROSAT-PSPC map. 
Note that, although this spike is cooler than the cluster, there is no
cooling flow: in fact, we derived a cooling time of $3.9 \times 10^{13}$ yr.
\\
In order to have an independent temperature estimate, we followed a more 
indirect way: we calculated the densities
of the King and of the Gaussian models at the center of the latter and 
imposed pressure equilibrium. In this case $n_e^{K} kT^{K}=n_e^{G} kT^{G}$,
where the indexes $K$ and $G$ refer to the two models. Given the fact that
$n_e^{G}$ depends on $kT^{G}$, we solve the equation iteratively,
finding $kT^{G}=1.12$ keV. 
\\
The pressure found at the center of the Gaussian results in 
$P=3.73\times 10^{-12}$ h$^{0.5}$ dyn cm$^{-2}$.
However, in the case where the Gaussian component has a temperature 
of $kT=2$ keV, the pressure is $P=5.90\times 10^{-12}$ h$^{0.5}$ dyn cm$^{-2}$:
in this case the pressure balance with the cluster is reached at a distance 
from the center of the order of $\sim 0.08$ \hmpc, corresponding to $1\sigma$
of the Gaussian. 
These estimates can only give an idea of the pressure inside 
the density spike in order to study the central radio source (see Sect. 5), 
but we have to take in mind the possibility that this entity is not at all 
at pressure equilibrium. 

\begin{figure}
\centering
\includegraphics[angle=90,width=\hsize]{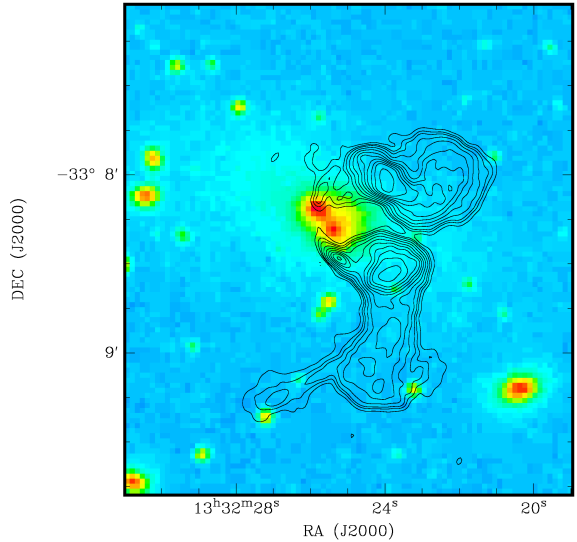}
\caption{Radio isocontours at a wavelenght of $20$ cm superimposed to the 
optical image taken from the Digital Sky Survey (DSS2). The resolution is
$4.2\times 3.7$ arcsec with a position angle of $61^o$. The map noise is 
$0.14$ mJy/beam and the first contour correspond to $0.45$ mJy/beam.
The other $n^{th}$ contours are spaced by a factor $2^n$. 
}
\label{fig:radiogal}
\end{figure}

\begin{figure}
\centering
\includegraphics[angle=0,width=\hsize]{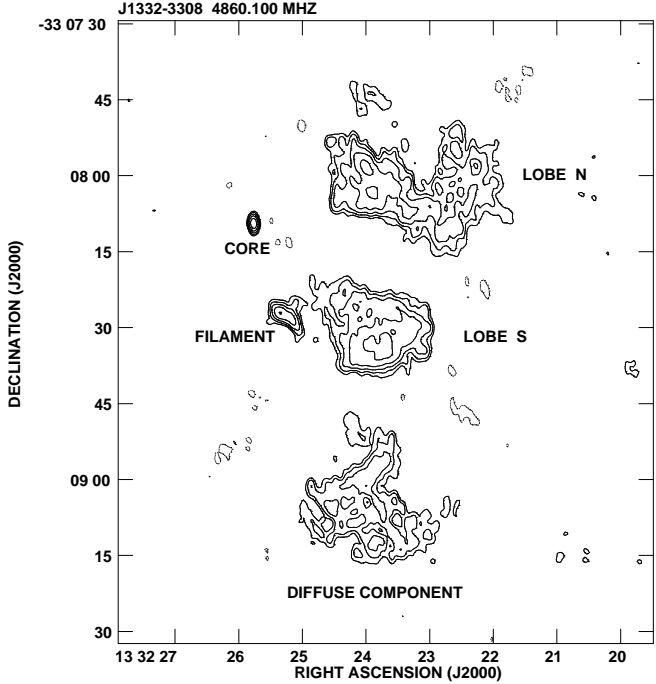}
\caption{Radio isocontours at a wavelenght of $6$ cm. The resolution is
$2.1\times 1.2$ arcsec with a position angle of $5^o$. The map noise is 
$46$ $\mu$Jy/beam and the first contour corresponds to $0.15$ mJy/beam.
The other $n^{th}$ contours are spaced by a factor $2^n$. 
}
\label{fig:radiogal2}
\end{figure}

%
\section{The radio galaxy J1332-3308}

At radio wavelengths A3560 is dominated by the extended 
radio galaxy J1332--3308, associated with the Dumb-bell galaxy
cited in Sect.2. 
This source is located approximately at the center of the peaked component
of the hot gas distribution (see Sect.3).
No other radio sources in the A3560 field are
associated with cluster galaxies.
\\
This radio galaxy was observed over a wide range of radio frequencies
with the Very Large Array (VLA, New Mexico, USA) and with 
the Australia Telescope Compact Array (ATCA, Narrabri, Australia) 
as part of a larger project addressing extended 
radio galaxies in the central region of the Shapley Concentration
(Venturi et al. in preparation). In this section we will present only 
those radio properties of this source relevant to the astrophysical
problems dealt with in the present paper.
\\ 
In Figure \ref{fig:radiogal} we show the 1.4 GHz VLA image of the 
radio galaxy  overlaid on the DSS--2 optical image.
As can be seen, 
J1332--3308 is a complex radio source, as is often found 
at the centre of galaxy clusters, with a total power of 
log $P$(W Hz$^{-1}$) = 24.41 at 1.4 GHz. 
This value can be considered the transition power between FRI 
and FRII radio galaxies
(Fanaroff \& Riley \cite{fr74}), and among the largest values found
for radio galaxies located at the centre of rich clusters (Burns 
\cite{burns90}; Ball et al. \cite{ball93}; 
Gregorini et al. \cite{gregorini94}).
The  origin of the radio emission is likely to  be associated with the
north-eastern component of the Dumb-bell galaxy.
\\
The higher resolution image at the wavelength of $6$ cm 
(Figure \ref{fig:radiogal2}) suggests that the overall 
morphology of the radio galaxy includes five different components: 
the nucleus of 
the radio emission, associated with the north-eastern optical nucleus 
of the optical counterpart, two ``fat" and extended lobes, labelled 
lobe N and lobe S, a small filament of radio emission located 
South of the radio core (pointing southwards), and a diffuse extended 
emission South of the lobe S. 
The connection between these components is unclear, and a detailed
discussion will be carried out in Venturi et al. (in preparation).
One possibility is that all components are coeval, and that the
complex morphology is the result of cluster weather, i.e. interaction with 
the surrounding intracluster gaseous medium (Burns \cite{burns98}), possibly
combined with orientation effects of the radio emission.
One alternative possibility is that J1332--3308 consists of one
``active" part, the nucleus and the two extended lobes N and S,
and of one ``relic" emission, the filament and the diffuse southern
structure.
\\
We estimated the equipartition parameters for this source, using
all the interferometric images available from 1.4 GHz to 8.6 GHz.
They are summarised in Table \ref{tab:radioparam}. 
The entries in the table are:
the component as described above (Col. 1); the spectral index 
in the range 1.4 GHz - 8.6 GHz, to be read in the sense 
S$\propto \nu^{-\alpha}$ (Col. 2); the equipartition magnetic
field (Col. 3); the internal pressure at equipartion (Col. 4).
We note that the equipartition parameters were computed taking into 
account cylindric geometry, a filling factor $\Phi$ = 1 and a ratio
between protons and electrons $K=1$.
\\
The equipartition parameters in the extended components (lobes and
filament) are those expected for the lobes and tails of
extended sources in clusters of galaxies (Feretti et al. \cite{feretti92}).
Overall, J1332--3308 is in pressure equilibrium with the external
medium (see Sect.3). 

\begin{table}
\caption[]{Physical parameters of J1332-3308}
\begin{flushleft}
\begin{tabular}{llll}
\hline\noalign{\smallskip}
Component & $\alpha_{1.4}^{8.6}$ & B$_{eq}$   &  P$_{eq}$ (10$^{-12}$) \\ 
~~~~~ & ~~~~~ &  $\mu$G & h$^{4/7}$ dyn cm$^2$    \\
\noalign{\smallskip}
\hline\noalign{\smallskip}
Core &  0.47 & 32.6 &  131.1 \\
Lobe N &  0.94 & 5.4 & 3.6 \\
Lobe S & 0.75 & 5.9 & 4.4 \\
Filament & 0.81 & 14.7 &  26.8 \\
Diffuse & 0.74 &  4.2&  2.2 \\     
\noalign{\smallskip}
\hline
\end{tabular}
\end{flushleft}
\label{tab:radioparam}
\end{table}

%
\section{Conclusions}

In this paper we studied A3560, a rich cluster (richness class 3)
at the southern periphery of the A3558 complex, a chain of interacting
clusters in the central part of the Shapley Concentration supercluster.
\\
From the ROSAT-PSPC map we found that the X-ray surface brightness 
distribution of A3560 is well 
described by two components, an elliptical King law and a more peaked and 
fainter structure, which has been modeled with a Gaussian. 
The main component, corresponding to the cluster, is elongated with the
major axis pointing toward the A3558 complex. The second component,
centered on the Dumb-bell galaxy which dominates the cluster, appears
significantly offset (by $\sim 0.15$ \hmpc) from the cluster X-ray 
centroid. However, the contribution of this component to the global luminosity
is only $\sim 2 \%$.
\\
From the Beppo-SAX observation we derived the radial temperature profile,
finding that the temperature is constant (at $kT \sim 3.7$ keV) up to
8 arcmin, corresponding to 0.3 \hmpc: for larger distances, the temperature
significantly drops to $kT \sim 1.7$ keV.
This drop questions the validity of the hydrostatic equilibrium hypothesis
for regions at distances $> 0.3$ \hmpc from the cluster center.
We also analyzed temperature maps, dividing the cluster into 4 sectors and
deriving the temperature profiles in each sector: we found that the temperature 
drop is significantly more sudden in sectors 1 and 2 (which point 
towards the A3558 complex), while it is smoother in the other sectors.
\\
From the VLA radio data, at 20 and 6 cm, we found a peculiar bright extended
radio source (J1332-3308), composed of a core (centered on the northern 
component of the Dumb-bell galaxy), two lobes, a ``filament" and a diffuse
component.
The filament is not aligned with any of the two lobes and seems to
point towards the diffuse component. 
\\
At a first look this source seems a Wide Angle Tail source, but the
filament and the diffuse component do not fit this scenario.
We suggest two possible interpretations of this:
\\
i) All components are related to the same nuclear activity: in this case
the filament and the diffuse emission are due to a strong interaction of the
radio source with the intracluster medium. 
This interaction could originate
from the offset position of the peaked X-ray component hosting the radio 
source with respect to the overall cluster (see Figure \ref{fig:chi2fit}). 
This offset can suggest a motion of the peaked component roughly along the
major axis of the cluster: the consequent ram pressure can be responsible
for the peculiarity of the radio source.  
\\
ii) The components are the result of an intermittency of the nuclear engine, 
following the model invoked for 3C~338 by Burns et al. (\cite{burns83}):
in this case the filament and the diffuse component are the remnants of
a previous activity of the radio source, while the core and the lobes
are the result of the present activity of the same source.
As in case i), also in this scenario a motion of the radio core in the 
North-South direction is required.
The only difference with respect to 3C~338 is that in our case all components 
have flatter spectral indexes, indicating that J1332--3308 is younger.
\\
Further investigations are needed to discriminate between these models: 
in particular, the interaction between the radio source and the cluster 
diffuse emission is a typical topic where the high resolution power of the 
Chandra satellite will be decisive.
\\
As a general conclusion, the elongation in the direction of the A3558 complex,
the offset of the peaked component with respect to the centroid of the cluster 
and its motion (suggested by the radio data) and the sudden drop in the
temperature profile seem to indicate that A3560 is a dynamically disturbed
cluster.
%
\begin{acknowledgements}
This research has made use of linearized event files produced
at the Beppo-SAX Science Data center.
This work has been partially supported by the Italian Space Agency grants
ASI-I-R-105-00 and ASI-I-R-037-01, and by the Italian Ministery (MIUR)
grant COFIN2001 ``Clusters and groups of galaxies: the interplay between 
dark and baryonic matter".
We thank the referee (dr. T.Tamura) for helpful comments. 
\end{acknowledgements}
%
%

%
\end{document}